\documentclass[a4paper,fleqn,usenatbib]{mnras}

\usepackage[T1]{fontenc}
\usepackage{ae,aecompl}
\usepackage{graphicx}	
\usepackage{amsmath}	
\usepackage{amssymb}	

\title[The Canarias Einstein Ring]{The Canarias Einstein Ring: a Newly Discovered Optical Einstein Ring}

\author[M.~Bettinelli et al.]{
M.~Bettinelli$^{1,2,3}$\thanks{E-mail: mbettine@iac.es, marghebettinelli@gmail.com},
M.~Simioni$^{1,2,3}$,
A.~Aparicio$^{2,1}$,
S.~L. Hidalgo$^{1,2}$,
S.~Cassisi$^{4,1}$,\newauthor
A.~R.~Walker$^{5}$,
G.~Piotto$^{3,6}$,
F.~Valdes$^{7}$
\\
$^{1}$Instituto de Astrof\`isica de Canarias, V\`{i}a L\`{a}ctea S/N, E-38200 La Laguna, Tenerife, Spain\\
$^{2}$Department of Astrophysics, University of La Laguna, E-38200 La Laguna, Tenerife, Canary Islands, Spain\\
$^{3}$Dipartimento di Fisica e Astronomia ``Galileo Galilei'', Universit\`{a} degli Studi di Padova,  Vicolo dell'Osservatorio 3, I-35122 Padova, Italy\\
$^{4}$INAF-Osservatorio Astronomico di Teramo, Via M. Maggini, I-64100 Teramo, Italy\\
$^{5}$Cerro Tololo Inter-American Observatory, National Optical Astronomy Observatory, Casilla 603, La Serena, Chile\\
$^{6}$INAF-Osservatorio Astronomico di Padova, Vicolo dell'Osservatorio 5, I-35122 Padova, Italy\\
$^{7}$National Optical Astronomy Observatory, P.O. Box 26732, Tucson, AZ 85719, USA
}

\date{Accepted 2016 May 12. Received 2016 April 18; in original form 2016 February 18.}

\pubyear{2016}

\begin{document}
\label{firstpage}
\pagerange{\pageref{firstpage}--\pageref{lastpage}}
\maketitle

\begin{abstract}
We report the discovery of an optical Einstein Ring in the Sculptor constellation, IAC J010127-334319, in the vicinity of the Sculptor Dwarf Spheroidal Galaxy. It is an almost complete ring ($\sim 300^{\circ}$) with a diameter of $\sim 4.5\, {\rm arcsec}$. 
The discovery was made serendipitously from inspecting Dark Energy Camera (DECam) archive imaging data.  Confirmation of the object nature has been obtained by deriving spectroscopic redshifts for both components, lens and source, from observations at the $10.4$ m Gran Telescopio CANARIAS (GTC) with the spectrograph OSIRIS. The lens, a massive early-type galaxy, has a redshift of ${\rm z}=0.581$ while the source is a starburst galaxy with redshift of ${\rm z}=1.165$.
The total enclosed mass that produces the lensing effect has been estimated to be ${\rm M_{tot}=(1.86 \pm 0.23) \,\cdot 10^{12}\, {\rm M_{\odot}}}$.

\end{abstract}


\section{Introduction}

Strongly lensed galaxies are very important in the study of  galaxy formation and evolution because they permit derivation of important physical parameters such as the total mass of the lensing object, without any assumption on the  dynamics. Cases in which the Einstein ring (ER) is almost complete and the central lensing galaxy isolated are rare; these permit constraining with great accuracy the enclosed mass within the projected Einstein radius $\Theta_{E}$ \citep{2001ApJ...547...50K}. \citet{1992MNRAS.259P..31M} predicted several $ 10^{6}$ optical ER to be detectable over the whole sky, down to a magnitude limit of $B=26$ and a lower limit for the enclosed mass of $M\sim5\cdot  10^{11}.\ M_{\odot}$. This notwithstanding, despite extensive surveys (see  for example \citealt{2008ApJ...682..964B,2013MNRAS.436.1040S}) only a few tens of complete or nearly complete optical ERs have been identified so far, and among these objects, only a few show a close similarity, in morphology and elongation of the ring, to the one we discuss in the present work.

The first ER to be discovered is the radio source MG1131+0456 \citep{1988Natur.333..537H}. 
\citet{1996MNRAS.278..139W} report the discovery of a partial ER ($\sim 170^{\circ}$) with $\Theta_{E}\sim 1.35\, {\rm arcsec}$; the background OII emitting galaxy at ${\rm z}=3.595$ is lensed by an elliptical massive galaxy at ${\rm z}=0.485$. This is the first known case in the literature of a ER discovered at optical wavelengths. \citet{2005A&A...436L..21C} discovered an almost complete ER ($\sim 260^{\circ}$) with $\Theta_{E}\sim 1.48\, {\rm arcsec}$ produced by a massive and isolated elliptical galaxy at ${\rm z}=0.986$. The source galaxy is a starburst at ${\rm z}=3.773$. Then, a similar ER to the one we report in this Letter, in morphology, but not in the physics of the source galaxy, a BX galaxy, is the so called "Cosmic Horseshoe" \citep{2007ApJ...671L...9B}; the ring extension is similar to the one we report here, ($\sim 300^{\circ}$), but the Einstein radius is double, $\Theta_{E}\sim 5\, {\rm arcsec}$; the lensing galaxy has a huge mass of $\sim {\rm M=5.4\, \times 10^{12}\, {\rm M_{\odot}}}$.  Other partial ER discovered recently are: the "Cosmic Eye" \citep{2007ApJ...654L..33S}, the "8 o'clock arc" \citep{2007ApJ...662L..51A} and the "Clone" \citep{2009ApJ...699.1242L}.

Here we report the discovery of IAC J010127-334319, an optical, almost complete ER, that we refer to as the "Canarias Einstein Ring", noticed as a peculiar object in DECam images.  No previous reference to the object has been found in the literature. 
Subsequently we observed it with OSIRIS@GTC for a spectroscopic confirmation of its nature. In this Letter we provide the first physical parameters of this system.
In the following discussion we assume a flat cosmology with $\Omega_{m}=0.3$, $\Omega_{\Lambda}=0.7$ and $H_{0}=70 \, {\rm km\, s^{-1}\, Mpc^{-1}}$. 

\begin{figure*}
\centering
\includegraphics[width=\textwidth]{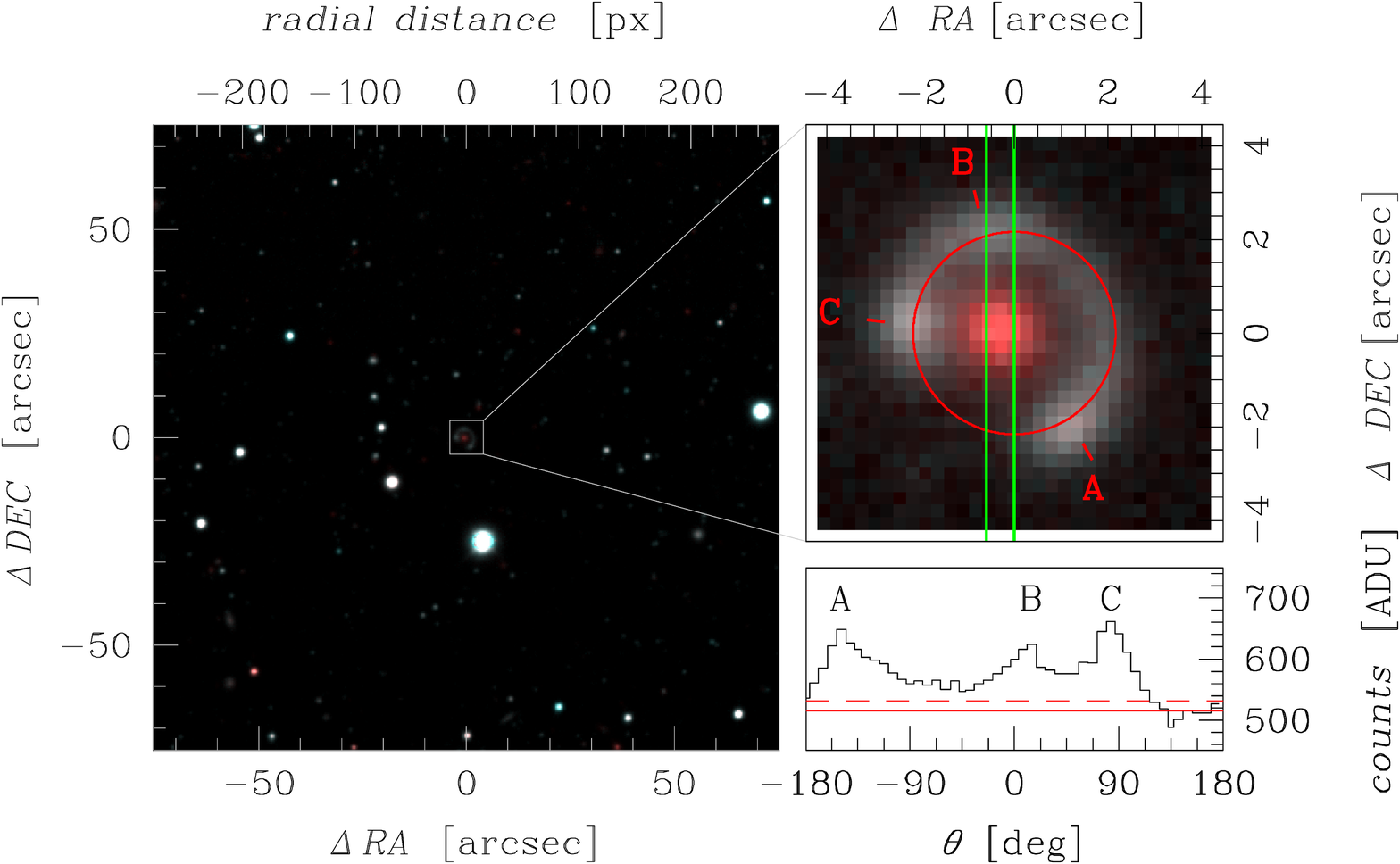}
\caption{Composite $g$, $r$ field of view of $2.5\, {\rm arcmin} \times 2.5\, {\rm arcmin}$ centered on the object (on the left), North is up, while East points left; a zoom of the object with overplotted the best fitting circle, the slit position and width are also plotted as green lines (right upper panel); counts from photometry along the best fit circle of the ring (lower right panel): the measured sky value is indicated by the red solid line, the $1\sigma$ value is indicated by the red dashed one.}
\label{fig:color}
\end{figure*}
\begin{table}
\footnotesize
\caption{List of parameters}
\begin{tabular}{ l r } 
\hline
Lens & {}\\ 
\hline \hline
Right ascension(J2000): & $01^{h}01^{m}27.83^{s}$\\
Declination(J2000): & $-33^{\circ}43^{'}19.68''$\\
Redshift: & ${0.581 \pm 0.001}$\\
Surface brightness Lens ($g$,$r$) ${\rm [mag\, arcsec^{-2}]}$: & $25.2,\, 22.2$ \\
Apparent magnitude ($g$,$r$) & $23.61,\,21.48$\\
Absolute magnitude ($g$,$r$) & $-21.05,\, -23.18$\\
\\
\hline
Ring & {}\\
\hline \hline
Redshift: & ${1.165 \pm 0.001}$\\
Einstein radius: & $2.16'' \pm 0.13$\\
Enclosed mass ${\rm [10^{12}\,M_{\odot}]}$: & $1.86 \pm 0.23$\\
Surface brightness A ($g$,$r$) ${\rm [mag\, arcsec^{-2}]}$: & $23.7,\, 22.9$ \\
Surface brightness B ($g$,$r$) ${\rm [mag\, arcsec^{-2}]}$: & $23.9,\, 23.2$ \\
Surface brightness C ($g$,$r$) ${\rm [mag\, arcsec^{-2}]}$: & $23.7,\, 23.0$ \\
Apparent magnitude ($g$,$r$) & $20.94,\,20.12$\\
\\
\hline
\end{tabular}
\label{tab:valori}
\end{table}

\section{DISCOVERY}

The serendipitous discovery of IAC J010127-334319 was made while performing photometry on stacked images, in $g$ and $r$ filters, taken with DECam \citep{2015AJ....150..150F} at the Blanco 4m telescope at the Cerro Tololo Inter-American Observatory (CTIO), reduced with the NOAO Community Pipeline \citep{2014ASPC..485..379V} and obtained from the NOAO Science Archive \citep{2002SPIE.4846..182S}. 
The total exposure time is $7680$ s in the $g$ filter and $5700$ s in the $r$ filter.
Figure~\ref{fig:color} shows the resulting color composite image: it is evident that two components with different colors are present. In particular the central one (the lens) appears redder than the second component (the lensed image of the source), which in turn appears as elongated all around the first. The ring is almost perfectly circular with an apparent radius of $8\,{\rm px}$ which translates to $2.16\, {\rm arcsec}$. Three peaks, A, B and C, are clearly visible (bottom right panel of Figure~\ref{fig:color}); they are located respectively at $-150\,{\rm deg}$, $14\,{\rm deg}$ and $83\,{\rm deg}$ from North counterclockwise.
In Table~\ref{tab:valori} all the derived parameters for the object are listed.
From the DECam photometry we estimated an apparent magnitude for the lens in both $g$ and $r$ bands of $g=23.61$, $r=21.40$.
The color $(g-r)> 2$ indicates that this galaxy is probably a luminous red galaxy \citep{2001AJ....122.2267E}. 
All the details about the photometric calibration will be given in a forthcoming paper (Bettinelli et al. in prep.). 

\section{FOLLOW-UP SPECTROSCOPY}
In order to confirm the lensing nature of this system we performed a spectroscopic follow-up at the $10.4$ m Gran Telescopio CANARIAS (GTC) on Roque de los Muchachos Observatory (La Palma, Spain) using the Optical System for Imaging and low-Intermediate-Resolution Integrated Spectroscopy  (OSIRIS) spectrograph \citep{1998Ap&SS.263..369C}. OSIRIS has a mosaic of two E2V CCD42-82 ($2048 \times 4096\, {\rm px}$). 
All the obtained spectra were registered on the second detector, which is the default for long-slit spectroscopy. We used a binning of $2 \times 2$ providing a pixel size of $0.254\, {\rm arcsec\, px^{-1}}$, and the grism R300B, which provides a spectral coverage of $4000-9000$ \AA\, and a nominal dispersion of $4.96$ \AA\,${\rm px^{-1}}$. The slit width was $0.6\, {\rm arcsec}$. 
Long-slit spectral observations were performed on 2015 December 2 in good seeing conditions of $\sim 0.8\, {\rm arcsec}$. The slit was placed along the N-S direction, in order to minimize the effects of atmospheric differential refraction at culmination.  The total exposure time was 3600 s divided into 6 exposures of 600 s each.  In each of the 6 exposures the two components, ring and lensing galaxy, have been detected and in particular their spectra were not overlapping. The position of the slit was such that the spectra obtained for the ring refers to peak B.

For the pre-reduction we have used the OSIRIS Offline Pipeline Software (OOPS); sky subtraction and flux calibration were performed using IRAF\footnote{IRAF is distributed by the National Optical Astronomy Observatory, which is operated by the Association of Universities for Research in Astronomy, Inc., under cooperative agreement with the National Science Foundation}.
We performed wavelength calibration using standard HgAr+Ne+Xe arc lamps; the resulting error on wavelength determination has been measured to be consistent with the above spectral resolution.
We corrected the extracted spectra for instrumental response using observations of the spectrophotometric standard star GD140, a white dwarf, obtained the same night. The fluxes of this standard star are available in \citet{1988ApJ...328..315M}.

\section{ANALYSIS AND DISCUSSION}

In order to derive the redshifts for the two components we noted the strong emission line in the source spectrum and the 4000 \AA\, Balmer discontinuity in the lens spectrum. This led us to choose template models for a starburst galaxy and a early-type galaxy respectively, as specified below. Following line identification, we determined redshifts. 

\begin{figure*}
\centering
\includegraphics[width=\textwidth]{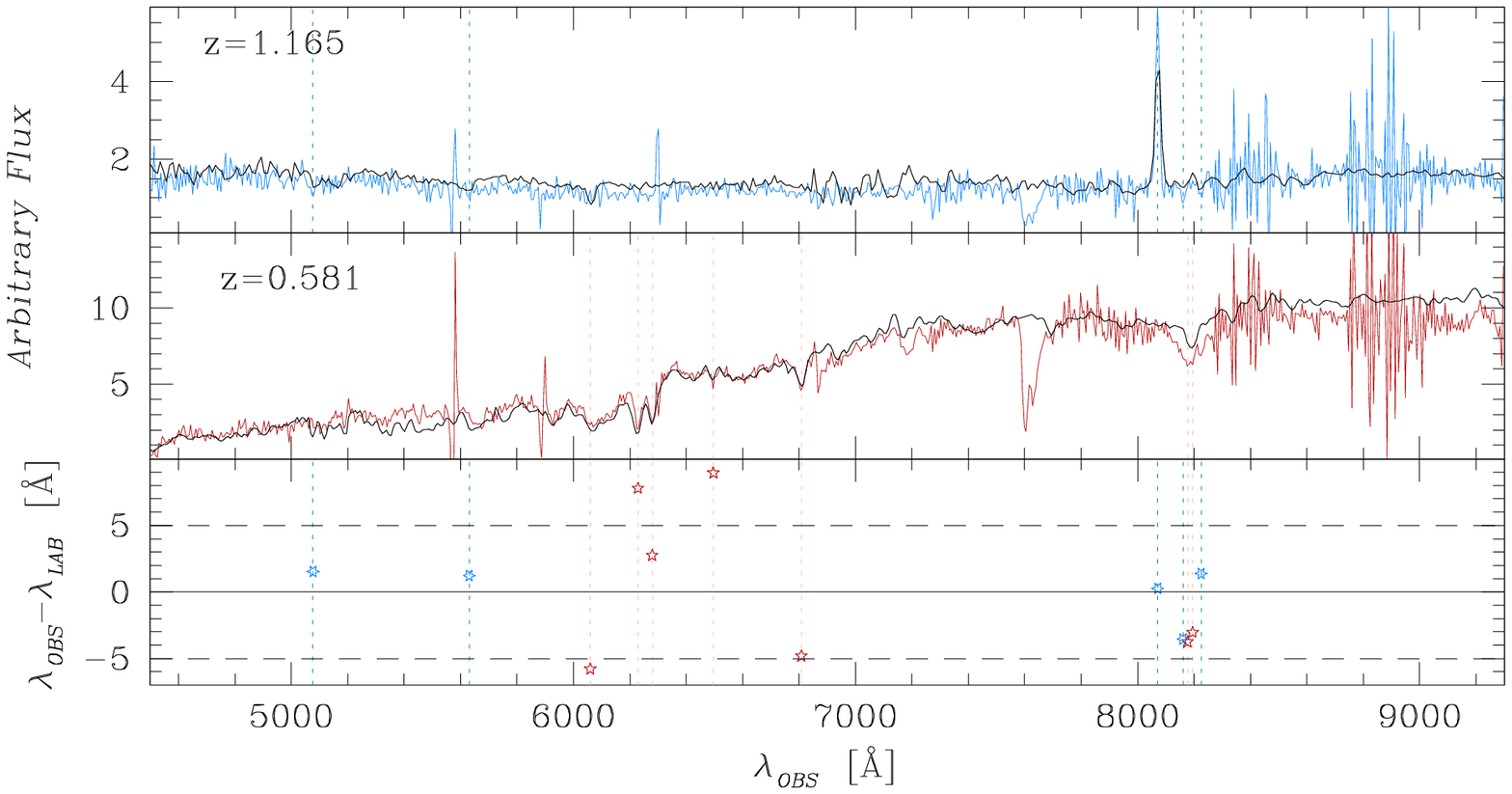}
\caption{Top panel: source galaxy spectrum (in blue) with overplotted a starburst template spectrum by \citet{1994ApJ...429..582C}. Middle panel: lens galaxy spectrum (in red) with overplotted an early-type galaxy template by \citet{1996ApJ...467...38K}. Bottom panel: measured wavelength displacement between observed and laboratory line position for the selected features (see text for details).}       
\label{fig:spec}
\end{figure*} 
\subsection{Lens}
Using the template spectra by \citet{1996ApJ...467...38K} results that the spectrum of the lens galaxy fits well the spectrum of a S0 galaxy (see Figure~\ref{fig:spec}), a typical early-type galaxy characterized by a large increase in flux from the UV part of the spectrum to the optical. The $4000$ \r{A} Balmer discontinuity at $\sim 6330$ \r{A} is noticeable. The redshift of the lens galaxy is ${\rm z}=0.581 \pm 0.001$ and it has been determined from the measurements of: ${\rm H\eta}\; \lambda3835.4$, Ca K $\lambda3933.7$, Ca H $\lambda3968.5$, ${\rm H\delta}\; \lambda4141.8$, G-band $\lambda4307.7$, Mg-b2 $\lambda5172.7$, Mg-b1 $\lambda5183.6$ (marked red features from left to right in Figure~\ref{fig:spec} middle panel). 

\subsection{Source}
For the source galaxy we used the template spectra by \citet{1994ApJ...429..582C} and we found that the spectrum best fitting our observed spectrum corresponds to a starburst galaxy in the case of \emph{clumpy scattering slab}, where it is assumed that clumped dust is located close to the source of radiation. In such circumstances, \citet{1994ApJ...429..582C} show that scattering into the line of sight dominates over absorption by the dust, providing a significant positive contribution to the emerging radiation.
This template spectrum fits well the strong \ion{O}{ii} $\lambda3727$  emission line. We also identified the following lines: \ion{Fe}{ii} $\lambda2344.0$, \ion{Fe}{ii} $\lambda2600.0$, \ion{H}{i}\,11 $\lambda3770.6$, \ion{O}{ii} $\lambda3727.3$,\ion{H}{i}\,10 $\lambda3797.9$ (marked blue features from left to right in Figure~\ref{fig:spec} upper panel). 
According to these features, we derived for the source galaxy a redshift of ${\rm z}=1.165 \pm 0.001$.
We note that the selected slit position enable us to extract only the portion of the spectrum corresponding to peak B (see Figure~\ref{fig:color}); this notwithstanding, the \ion{O}{ii} emission coming from the opposite side of the ring can be easily noted in our spectra.

\begin{figure}
\centering
\includegraphics[width=\columnwidth]{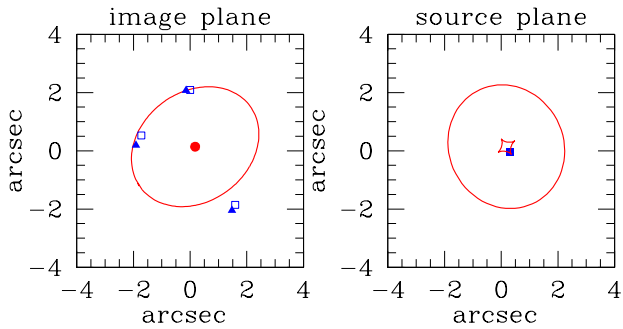}
\caption{Best fitting SIE model obtained with the {\ttfamily gravlens/lensmodel} software. On the left (image plane): the source images positions are plotted as blue triangles, the fitted position recovered by the software as blue squares, the red central dot represents the position of the lensing galaxy and the red curve is the critical curve. On the right (source plane): the blue square represent the calculated position of the source; the caustics are shown in red.}       
\label{fig:mass} 
\end{figure} 

\subsection{Enclosed Mass Derivation}
The strong circular symmetry of our object (see Figure~\ref{fig:color}) suggests that it can be approximated to the case of a circularly symmetric lens,  with source and lens in the line of sight.
Under these assumptions, for an arbitrary mass profile ${\rm M(\Theta)}$, (i.e. without assuming any particular model for the potential), we can apply the following relation \citep{1996astro.ph..6001N} and solve it for the mass.

\begin{equation}
\Theta^{2}_{\rm E}\, =\, \frac{4{\rm G}}{\rm c^{2}}\, {\rm M}(\Theta)\, \frac{d_{\rm LS}}{d_{\rm L}d_{\rm S}}
\end{equation}

Here $\Theta_{\rm E}$ is the Einstein radius in radians; ${\rm M(\Theta)}$ is the mass enclosed within the Einstein radius; $d_{\rm LS}$, $d_{\rm L}$, $d_{\rm S}$ are the angular diameter distances respectively of source-lens, lens-observer and source-observer. These last quantities are related to the relative comoving distances and, in general, this relation depends on the assumed curvature of the Universe \citep{1999astro.ph..5116H}. In our case $\Omega_{K}=0$ has been assumed and the resulting angular diameter distances are $d_{\rm L}=951\,{\rm h^{-1}Mpc}$, $d_{\rm S}=1192\,{\rm h^{-1}Mpc}$ and $d_{\rm LS}=498\,{\rm h^{-1}Mpc}$.
We calculated a total mass ${\rm M_{tot}=(1.86 \pm 0.23) \,\cdot 10^{12}\, {\rm M_{\odot}}}$ where the error on the mass ($12\%$) is overwhelmingly due to the measurement error in the determination of the Einstein radius, that we have
estimated to be $0.5\, {\rm px}$ which corresponds to $0.135\, {\rm arcsec}$.
The error on the redshift derives from the error estimated on the wavelength calibration which is $\sim 5$ \AA. This value is consistent with the spectral resolution ($4.96$ \AA\, ${\rm px^{-1}}$) of the grating R300B that we used.

Under the assumption of a singular isothermal sphere (SIS) it is possible to give an estimate of the magnification of the ring: $\mu=4\Theta_{E}/\delta\Theta_{s}$, where $\delta\Theta_{s}$ is the source size. From \citet{2011ApJ...735L..19N} the average size of a starburst galaxy in our redshift range is $\sim 2$ Kpc which corresponds to $0.24\, {\rm arcsec}$. The derived magnification is $\sim 36$.

We determined also the mass-to-light ratio of the lens in the ${\rm g}$ band; a K-correction of 2.1 has been derived for the lens using the NED calculator \citep{2012MNRAS.419.1727C}. The resulting ratio is ${\rm M_{tot}/L} \sim 58\, {\rm M_{\odot}/L_{\odot}}$. 

The former mass estimate can be improved by applying to the system a singular isothermal ellipsoid (SIE) model using the {\ttfamily gravlens/lensmodel} software \citep{2001astro.ph..2340K}. This software allows to fit a SIE model using only the image positions and fluxes. The obtained best-fit model is plotted in Figure~\ref{fig:mass}.
The solutions have been derived for pointlike sources; the ring shape is due to the fact that the source should actually be more extended with respect to the caustics than is shown here.
The best fit ellipticity is $0.2$ calculated as $1-q$ where $q$ is the axis ratio; the associated position angle is -57 deg, angle measured from North to East. 
The best fit $\chi^{2}$ value is 6.12, calculated setting to 0 the weights relative to image fluxes.
The derived Einstein radius is $\Theta_{E}=2.38\, {\rm arcsec}$, which translates in an enclosed mass of $\sim 2.26\,\cdot 10^{12}\, {\rm M_{\odot}}$, hence in excellent agreement with our previous estimate.

\section{Conclusions}

We report the discovery of an almost complete ($\sim 300^{\circ}$) circular optical Einstein ring in the constellation of Sculptor. 
The gravitational lens is a massive luminous red galaxy at $z=0.581$. The source galaxy is a starburst at redshift $z=1.165$; its spectrum is dominated by a strong \ion{O}{ii} emission line. Using these redshift determinations and the Einstein radius $\Theta_{E}=2.16\, {\rm arcsec}$ we calculated the total enclosed mass that produced the lensing effect: ${\rm M_{tot}=(1.86 \pm 0.23) \,\cdot 10^{12}\, {\rm M_{\odot}}}$.

All the parameters we determined for IAC J010127-334319 are listed in Table~\ref{tab:valori}.

\bigskip 

\section*{Acknowledgements}

The authors thank the anonymous referee for the constructive comments that significantly improved the manuscript. 
The authors are grateful to all the GTC staff and in particular to Dr. A. Cabrera-Lavers for his support in refining the spectroscopical observations during the Phase-2. 
The authors also thank Dr. J. Falc\'{o}n-Barroso for the helpful discussion.

This Letter is based on observations made with the GTC telescope, in the Spanish Observatorio del Roque de los Muchachos of the Instituto de Astrof\'isica de Canarias, under Director's Discretionary Time.

This project used data obtained with the Dark Energy Camera (DECam), which was constructed by the Dark Energy Survey (DES) collaboration.
Funding for the DES Projects has been provided by 
the DOE and NSF (USA), 
MISE (Spain), 
STFC (UK), HEFCE (UK), 
NCSA (UIUC), 
KICP (U. Chicago), 
CCAPP (Ohio State), 
MIFPA (Texas A\&M), 
CNPQ, FAPERJ, FINEP (Brazil), 
MINECO (Spain), 
DFG (Germany) 
and the collaborating institutions in the Dark Energy Survey, which are 
Argonne Lab, 
UC Santa Cruz, 
University of Cambridge, 
CIEMAT-Madrid, 
University of Chicago, 
University College London, 
DES-Brazil Consortium, 
University of Edinburgh, 
ETH Z{\"u}rich, 
Fermilab, 
University of Illinois, 
ICE (IEEC-CSIC), 
IFAE Barcelona, 
Lawrence Berkeley Lab, 
LMU M{\"u}nchen and the associated Excellence Cluster Universe, 
University of Michigan, 
NOAO, 
University of Nottingham, 
Ohio State University, 
University of Pennsylvania, 
University of Portsmouth, 
SLAC National Lab, 
Stanford University, 
University of Sussex, 
and Texas A\&M University.

M.B., M.S., A.A., S.L.H., S.C. and G.P. acknowledge support from the Spanish Ministry of Economy and Competitiveness (MINECO) under grant AYA2013-42781.

This research has made use of the NASA/IPAC Extragalactic Database (NED) which is operated by the Jet Propulsion Laboratory, California Institute of Technology, under contract with the National Aeronautics and Space Administration.

\bsp	
\label{lastpage}
\end{document}